\documentclass[dvips]{article}
\usepackage{epsfig}

\def\ageq{\vcenter{\vbox{\hbox{$\buildrel > \over 
\sim$}}}}

\begin{document}
\textwidth=12.7cm
\textheight=18.4cm
\begin{center}
\section*{DIQUARKS AS EFFECTIVE PARTICLES IN HARD EXCLUSIVE
SCATTERING
\footnote{Talk given by W. Schweiger
at the \lq\lq International Conference on Nuclear and
Particle Physics with CEBAF at Jefferson Lab\rq\rq, Dubrovnik,
Croatia, Nov. 1998.
}}
\vspace{0.5 cm} CAROLA F. BERGER, BERNHARD LECHNER,\\ and WOLFGANG
SCHWEIGER\\
\vspace{0.5 cm} {\it Institute of Theoretical Physics, University of
Graz \\A-8010 Graz, Universit\"atsplatz 5, AUSTRIA\\ email:
wolfgang.schweiger@kfunigraz.ac.at}\\
\end{center}
\vspace{0.3 cm} 
In the context of hard hadronic reactions diquarks are a useful
phenomenological device to model non-perturbative effects still
observable in the kinematic range accessible by present-day
experiments.  In the following we present diquark-model
predictions for
$\gamma\gamma \rightarrow p \bar{p}$ and $\Lambda \bar{\Lambda}$.
We also sketch how the (pure quark) hard-scattering formalism for
exclusive reactions involving baryons can be reformulated in terms
of quarks and diquarks. As an application of these
considerations we analyze the magnetic proton form factor with
regard to its quark-diquark content.\newline


\noindent Keywords: perturbative QCD, diquarks, hard hadronic
processes, two-gamma reactions, proton magnetic form factor\hfill\\

In a series of papers \cite{JKSS93,Kro93,Kro96,KSPS96} (and
references therein) a systematic study of hard exclusive reactions
has been attempted within a model based on perturbative
QCD in  which baryons, however, are treated as quark-diquark rather
than three-quark systems.  The processes which have been treated 
in a consistent way as yet include baryon form factors in the space-
\cite{JKSS93} and time-like region \cite{Kro93}, real and virtual
Compton scattering \cite{Kro96}, two-photon annihilation into
proton-antiproton \cite{Kro93}, the charmonium decay $\eta_{{\rm c}}
\rightarrow p \bar{p}$
and
photoproduction of the
\hbox{$K^{+}$-$\Lambda$} final state \cite{KSPS96}. Like the usual
hard-scattering formalism (HSF) for exclusive hadronic reactions
\cite{BL89} the diquark model is based on factorization of short-
and
long-distance dynamics; a hadronic amplitude is expressed as a
convolution of a hard-scattering amplitude, calculable within
perturbative QCD, with distribution amplitudes (DAs) which contain
the (non-perturbative) bound-state dynamics of the hadronic
constituents. The introduction of diquarks is, above all, motivated
by the requirement to extend the HSF from (asymptotically) large
down to intermediate momentum transfers ($p_{\perp}^2 \,
\ageq \, 4\,  \hbox{GeV}^2$). This is the momentum-transfer region
where some experimental data exist, but where still
persisting non-perturbative effects, observable, e.g., as scaling
violations or violation of hadronic helicity conservation, prevent
the pure quark HSF to become fully operational. Diquarks may thus be
considered as an effective way to cope with such effects. 
\begin{figure}[t!]
\begin{center}
\epsfig{file=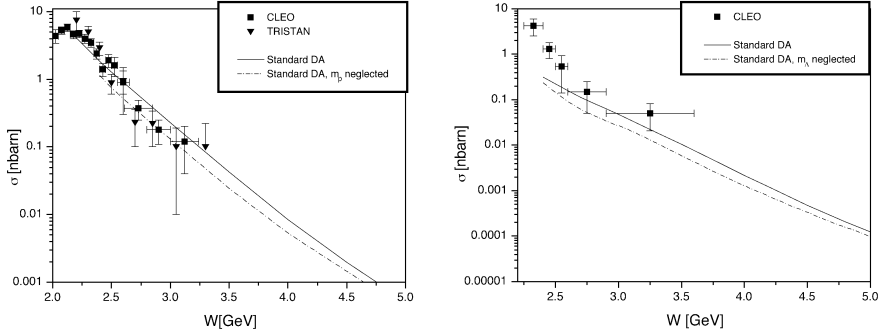,width=12.0cm,clip=}
\end{center}
\caption{Integrated $\gamma \gamma \rightarrow p \bar{p}$ (left) and 
$\gamma \gamma \rightarrow \Lambda \bar{\Lambda}$ (right) cross
sections  ($\vert \cos(\theta_{\rm cm})\vert \leq 0.6 $) vs. $W =
\protect\sqrt{s}$. Solid (dash-dotted) line: predictions obtained
with the standard parameterization of the diquark model (cf.
Ref.~\protect\cite{JKSS93}) with (without) mass corrections. Data
are taken from Refs. \protect\cite{CLEO94, VENUS97, CLEO97}.
\label{fig1}}
\end{figure}

The model, as applied in Refs.~\cite{JKSS93,Kro93,Kro96,KSPS96},
comprises scalar (S) as well as axial-vector (V) diquarks.
V-diquarks are important if one wants to describe spin observables
which require the flip of baryonic helicities. For the Feynman rules
of electromagnetically and strongly interacting diquarks, as well as
for the choice of the quark-diquark
distribution amplitudes of octet
baryons we refer to Ref.~\cite{JKSS93}. Here it is only important to
mention that the composite nature of diquarks is taken into account
by multiplying each of the Feynman diagrams entering the hard
scattering amplitude with diquark form factors. These are
parameterized by  multipole functions  with the power chosen in
such a way that in the limit $p_{\perp}
\rightarrow \infty$ the scaling behavior of the pure quark HSF is
recovered.

We want to present here a very recent application of the diquark 
model concerning the class of reactions $\gamma \gamma 
\rightarrow B \bar{B}$, where $B$ represents an octet baryon.
In contrast to foregoing work, we have now considered these
processes within the full model including also vector-diquarks.
Furthermore,  baryon-mass effects are taken into account in a
rigorous way by means of a systematic expansion in the parameter
(baryon mass/photon energy). With the same set of model parameters
as in  Refs.~\cite{JKSS93,Kro93,Kro96,KSPS96} we find that the
integrated cross-section data (available only) for the
$p$-$\bar{p}$ and the
$\Lambda$-$\bar{\Lambda}$ channel are very well reproduced (cf.
Fig.~1). By comparing the solid and the dash-dotted line it can also
be observed that in the few-GeV range baryon-mass effects are still
sizable. For details of the calculation and results for other
octet-baryon channels we refer to Ref.~\cite{Be97}.

\begin{table}[t!]
\caption{ Decomposition of the proton magnetic form factor into
diquark contributions for the proton DA proposed by Chernyak
et al.~\protect\cite{COZ89}. $S^{I}[q_1 q_2]$ and 
$V^{I}_{h}[q_1 q_2]$ denote scalar and vector diquarks (with isospin
$I$ and helicity
$h$), respectively, consisting of quarks $q_1$ and $q_2$. The
various diquark contributions are further decomposed into a 3- and
4-point part, $G_{\rm M}^{3, {\rm p}}$ and $G_{\rm M}^{4, {\rm p}}$,
depending on whether one or two gauge bosons couple to the diquark. 
The constant $C= 1.266$ is chosen in such a way that the largest
contribution, i.e. the $S^0[ud]\rightarrow S^0[ud]$ transition,
becomes~1.
\label{tab1}}
\vspace{0.5 cm}
\begin{center}
\begin{tabular}{|c|r|r|r|}\hline
\hline Transition & $C^{-1}Q^4 G_{\rm M}^{3, {\rm p}}$ & $C^{-1}Q^4
G_{\rm M}^{4, {\rm p}}$&{\rm Sum}\rule[-5pt]{0pt}{15pt}\\
\hline
$V_0^1[ud]\rightarrow V_0^1[ud]$ & 0.016 & 0.003 & 0.019
\rule[-4pt]{0pt}{13pt}\\ \hline
$V_0^1[ud]\leftrightarrow V_0^0[ud]$ & 0 & 0.004 & 0.004
\rule[-4pt]{0pt}{13pt}\\ \hline
$V_0^1[ud]\leftrightarrow S^1[ud]$ & 0 & 0 & 0
\rule[-4pt]{0pt}{13pt}\\ \hline
$V_0^1[ud]\leftrightarrow S^0[ud]$ & 0 & -0.063 & -0.063
\rule[-4pt]{0pt}{13pt}\\ \hline
$V_0^0[ud]\rightarrow V_0^0[ud]$ & 0.001 & 0 & 0.001
\rule[-4pt]{0pt}{13pt}\\ \hline
$V_0^0[ud]\leftrightarrow S^1[ud]$ & 0 & -0.002 & -0.002
\rule[-4pt]{0pt}{13pt}\\ \hline
$V_0^0[ud]\leftrightarrow S^0[ud]$ & 0 & -0.011 & -0.011
\rule[-4pt]{0pt}{13pt}\\ \hline
$S^1[ud]\rightarrow S^1[ud]$ & $\approx 0 $& $\approx 0$ & 0.001
\rule[-4pt]{0pt}{13pt}\\ \hline
$S^1[ud]\leftrightarrow S^0[ud]$ & 0 & -0.002 & -0.002
\rule[-4pt]{0pt}{13pt}\\ \hline
$S^0[ud]\rightarrow S^0[ud]$ & 1.000 & -0.005 & 0.995
\rule[-4pt]{0pt}{13pt}\\ \hline
$V^1_1[ud]\rightarrow V^1_1[ud]$ & 0 & -0.002 & -0.002
\rule[-4pt]{0pt}{13pt}\\ \hline
$V^1_1[ud]\leftrightarrow V^0_1[ud]$ & 0 & 0.036 & 0.036
\rule[-4pt]{0pt}{13pt}\\ \hline
$V^0_1[ud]\rightarrow V^0_1[ud]$ & 0 & 0.007 & 0.007 
\rule[-4pt]{0pt}{13pt}\\ \hline
$V_0^1[uu]\rightarrow V_0^1[uu]$ & -0.016 & 0.027 & 0.012
\rule[-4pt]{0pt}{13pt}\\ \hline
$V^1_0[uu]\leftrightarrow S^1[uu]$ & 0 & -0.003 &-0.003
\rule[-4pt]{0pt}{13pt}\\ \hline
$S^1[uu]\rightarrow S^1[uu]$ & $\approx 0 $& 0.005 & 0.004
\rule[-4pt]{0pt}{13pt}\\ \hline
$V^1_1[uu]\rightarrow V^1_1[uu]$ & 0 & -0.013 & -0.013
\rule[-4pt]{0pt}{13pt}\\ \hline
\multicolumn{3}{|c|}{Total}& 0.985 \rule[-4pt]{0pt}{13pt}\\ \hline
\end{tabular}
\end{center}
\end{table}
 
As the applications mentioned above demonstrate, diquarks are
obviously a very useful phenomenological concept (not only) in the
field of hard hadronic processes. Physically speaking, diquarks
represent effective particles which describe strong quark-quark
correlations in baryonic wave functions. Within the pure quark HSF
such correlations seem indeed necessary to obtain reasonable
results, even for the simplest exclusive observables such as the
nucleon magnetic form factors~\cite{COZ89}.
A more formal
justification of diquarks can be obtained by observing that the
diquark model should evolve into the pure quark HSF in the limit of
asymptotically large momentum transfers. This suggests a
reformulation of the pure quark HSF in terms of quark and diquark
degrees of freedom. Two obvious constraints for this reformulation
are that the leading order hard-scattering amplitude on the
quark-diquark level should also consist only of tree graphs
(like in the pure quark HSF) and that the result of this
reformulation should be
independent of the choice of the two quarks which are grouped to a
diquark. It has been proved in Ref.~\cite{Le97} that a
reformulation of the pure quark HSF fulfilling both constraints is
indeed possible. If we employ this
reformulation to analyze the proton magnetic form factor with
respect to its diquark content, we find the isospin $0$ scalar
$S[ud]$ diquark to provide the by far most important contribution
(cf. Tab.~1). This is not only the case for the proton DA proposed
by Chernyak et al. but holds also for other DA models.  

The reformulation of the pure quark HSF in terms of quarks and
diquarks requires to study the general Lorentz structure of
two-quark subgraphs to obtain the Lorentz covariants and
corresponding (Lorentz-invariant) vertex functions of the various
gauge-boson diquark vertices. This gives valuable clues how
gauge-boson diquark vertices and corresponding form factors could
be improved in the naive diquark model. However, in order to arrive
at an effective model in the sense that it reproduces the results
of the pure quark HSF (and not only the scaling behavior) in the
limit of asymptotically large momentum transfers one should take
these vertices literally and use the vertex-function results as
asymptotic constraints for the parameterization of the diquark form
factors. A corresponding program is presently carried out.

\end{document}